\documentclass[twocolumn,prb,showpacs,superscriptaddress]{revtex4-1}

\usepackage{graphicx}
\usepackage{bm}
\usepackage{amssymb, amsmath, newtxmath, newtxtext}
\usepackage{color}

\newcommand{\ve}[1]{\boldsymbol{#1}}

\begin{document}

\title{Deviation from the Fermi-Liquid Transport Behavior in the Vicinity of a Van Hove Singularity}
\author{Franti\v{s}ek Herman}
\affiliation{Institute for Theoretical Physics, ETH Zurich, 8093 Zurich, Switzerland}
\author{Jonathan Buhmann}
\affiliation{Institute for Theoretical Physics, ETH Zurich, 8093 Zurich, Switzerland}
\author{Mark H Fischer}
\affiliation{Department of Physics, University of Zurich, 8057 Zurich, Switzerland}
\affiliation{Institute for Theoretical Physics, ETH Zurich, 8093 Zurich, Switzerland}
\author{Manfred Sigrist}
\affiliation{Institute for Theoretical Physics, ETH Zurich, 8093 Zurich, Switzerland}

\begin{abstract}
Recent experiments revealed non-Fermi-liquid resistivity in the unconventional superconductor Sr$_{2}$RuO$_{4}$ when strain pushes one of the Fermi surfaces close to a van Hove singularity.
The origin of this behavior and whether it can be understood from a picture of well defined quasiparticles is unclear.
We employ a Boltzmann transport analysis beyond the single relaxation-time approximation based on a single band which undergoes a Lifshitz transition, where the Fermi surface crosses a van Hove singularity, either due to uni-axial or epitaxial strain.
First analytically investigating impurity scattering, we clarify the role of the diverging density of states together with the locally flat band at the point of the Lifshitz transition. 
Additionally including electron-electron scattering numerically, we find good qualitative agreement with resistivity measurements on uni-axially strained Sr$_{2}$RuO$_{4}$, including the temperature scaling and the temperature dependence of the resistivity peak.
Our results imply that even close to the Lifshitz transition, a description starting from well-defined quasiparticles holds. 
To test the validity of Boltzmann transport theory near a van Hove singularity, we provide further experimentally accessible parameters, such as thermal transport, the Seebeck coefficient, and Hall resistivity and compare different strain scenarios.
\end{abstract}
\maketitle

\section{Introduction}

Fermi liquid theory, which establishes a one-to-one mapping of electrons to well-defined quasiparticles, is a basis of our understanding of metals. Its validity as well as its breakdown are often characterized through transport properties. In particular, the quadratic temperature dependence of the resistivity $\rho = \rho_0 + AT^2$ due to electron-electron scattering and the linear-in-$T$ Seebeck coefficient $Q$ are hallmarks of a Fermi liquid. Despite the theory's great success in describing most metals, some classes of systems are known to violate these expectations, most notably interacting fermions in one dimension and systems close to a quantum critical point.

Bringing the Fermi surface close to a van Hove singularity (vHS), i.e., a point of diverging density of states, provides another example, where non-Fermi-liquid behavior can be observed. 
Whether such behavior is associated with a breakdown of Fermi liquid theory and the disappearance of well-defined quasiparticles is, however, not well established. 
Indeed,  \textcite{buhmann:2013b} showed how non-generic transport behavior can be observed within a picture of well-defined quasiparticles subject to electron-electron (Umklapp) scattering close to a vHS. 

Experimentally, modifying the Fermi energy through doping and thus moving the Fermi surface close to a vHS, on the one hand, is straight-forward. However, this introduces disorder into the system, making comprehensive transport studies impossible. 
Fermi surface engineering through tensile or compressive strain, on the other hand, provides a non-invasive method for tuning the electronic structure of a material~\cite{hicks:2014, overcash:1981, watlington:1977, welp:1992}. 
Indeed, recent experiments have demonstrated both routes on the single-layer perovskite Sr$_2$RuO$_4$, which at low temperature exhibits almost perfect Fermi-liquid behavior below $T\approx 50$K~\cite{maeno:1997} before entering a superconducting state at $T_c\approx 1.5$K. 
Interestingly, the so-called $\gamma$ band stemming mostly from Ru $d_{xy}$ orbitals, nearly touches the Brillouin zone (BZ) boundary, where the vHS is located ~\cite{bergemann:2003} and is, therefore, most interesting in connection with Fermi surface tuning. 

\begin{figure}[b!]
	\includegraphics[width=8.7 cm]{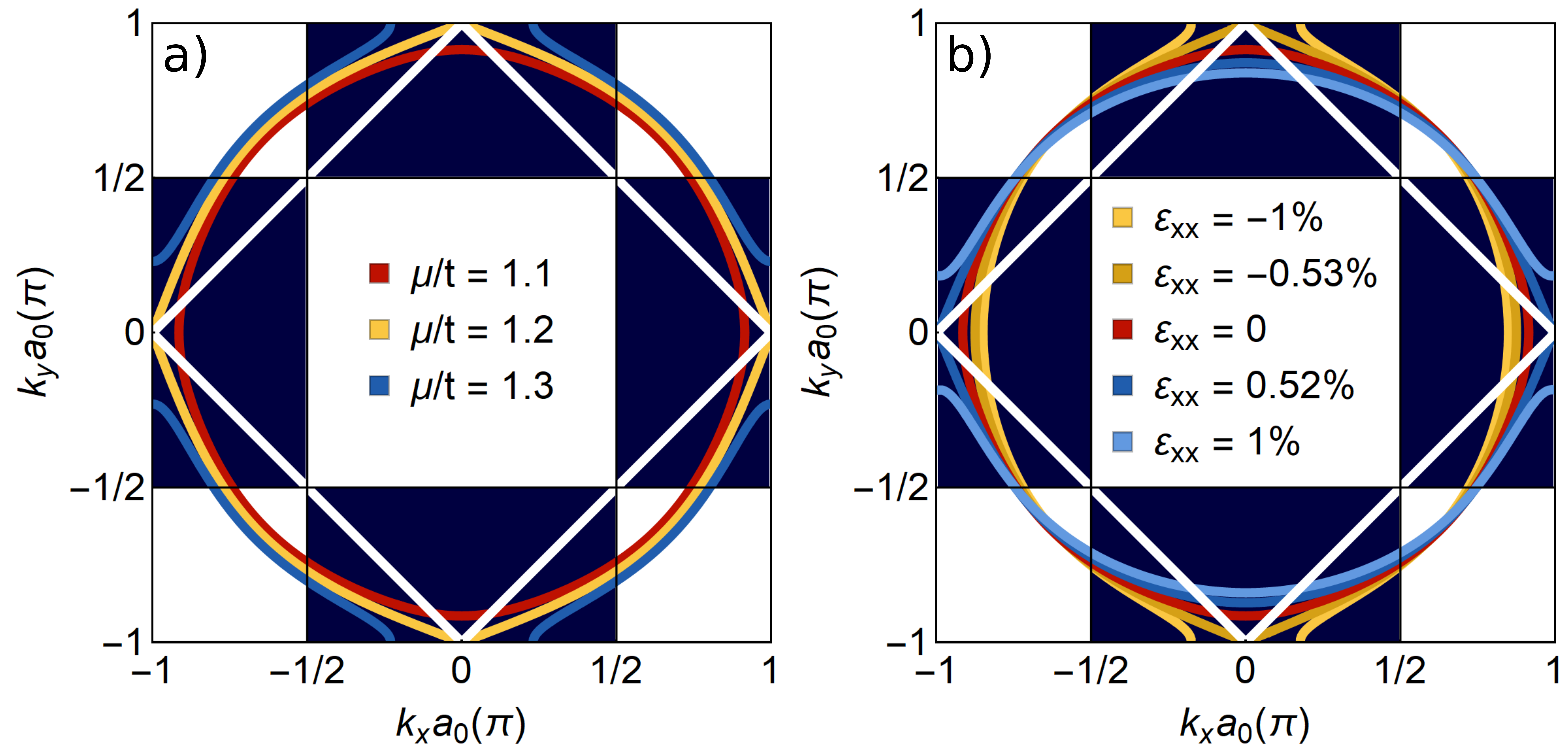}
	\caption{
	Fermi surface tuned through the vHS at $(\pi, 0)$ or $(0, \pi)$ by a) adjusting the chemical potential (`doping') and b) applying uniaxial stress. The dark areas [white diamond] denote regions, where Umklapp scattering by $(2\pi, 0)$ or $(0, 2\pi)$ [$(2\pi, 2\pi)$] are possible. 
	}
	\label{Fig:DOS_Gamma_Band_Umklapp}
\end{figure}

\textcite{barber:2018} showed that uni-axial stress can be used as a tuning knob for the normal state resistivity.
Here, DFT calculations indicate that the $\gamma$ Fermi surface undergoes a Lifshitz transition~\cite{lifshitz:1960} [see Fig.~\ref{Fig:DOS_Gamma_Band_Umklapp} b)] at a critical stress that in resistivity measurements coincides with a pronounced peak at low temperatures and $T$-linear scaling above.
An alternative route for Fermi surface engineering was demonstrated by \textcite{burganov:2016}, who epitaxially grew thin films on lattice-mismatched substrates. The resulting strain leads to a redistribution of electrons within the $t_{2g}$ manifold effectively `doping' the $\gamma$ band. This effective doping can induce a Lifshitz transition as well with associated crossing of a vHS as indicated in Fig.~\ref{Fig:DOS_Gamma_Band_Umklapp} a) and deviations from $T^2$ behavior were observed in the resistivity on the samples close to the transition. Having access to the exposed surface, this setup additionally allows for characterization of the electronic structure using STM and ARPES.

While previous work on the effect of a Lifshitz transition on the transport properties~\cite{varlamov:1989, hlubina:1996, markiewicz:1997} used approximations to the electron-electron scattering within a Green's function method, or through weighted scattering probabilities, our approach is based on an analysis (both analytic and numerical) of the general transport coefficients within a Boltzmann-equation approach including effects of Umklapp scattering.
The aim of our work is then twofold: First, our numerical approach taking both impurity scattering and interaction effects including Umklapp scattering into account allows for a careful comparison to existing experimental results of electrical transport upon Fermi surface tuning. The observed agreement with experiment establishes that well-defined quasi-particles are indeed capable of capturing all observed trends. 
Second, we predict further transport signatures near the vHS, such as thermal transport, Seebeck coefficient, Hall effect, a violation of the Wiedemann-Franz law, and the Kadowaki-Woods ratio. 
Here, we compare the two scenarios of crossing the vHS at all four boundaries of the Brillouin zone and the Lifshitz transition only at two points in the direction parallel or orthogonal to an applied electrical field, see Figs.~\ref{Fig:DOS_Gamma_Band_Umklapp} a) and b). These predictions allow for additional comparison of the vHS scenario within a picture of well-defined quasiparticles to experiment.

In the following we first introduce the general Boltzmann formalism and model-specific details for both types of scatterings. An analytical discussion of electrical transport for impurity scattering which is mainly relevant for the low temperature regime explains some of the main features connected with the crossing of the Fermi surface through the vHS. Then we will turn to the numerical discussion and examine the different transport properties showing how deviations from standard Fermi liquid behavior occurs in connection with the Lifshitz transitions within both scenarios of Fermi surface tuning. 
As we can reproduce many of the experimental findings within our calculation we conclude that the non-Fermi liquid behavior can be accounted for within the picture of intact quasi-particles. Many of the calculated quantities have not been investigated on real systems so far, such that our results could be further tested in experiment.

\section{Boltzmann approach}
\label{sec:boltzmann}
We investigate general transport properties by solving the Boltzmann transport equation
\begin{multline}\label{eq:BE_full}
  \big(\partial_t + \dot{\ve{r}}\!\cdot\!{\ve{\nabla}}_{\ve{r}}+ \dot{{\ve{p}}}\!\cdot\!{\ve{\nabla}}_{\ve{p}}\big)f({\ve{u}}) = \big[\partial_t f({\ve{u}})\big]_{\rm imp} + \big[\partial_t f({\ve{u}})\big]_{\rm el-el},
\end{multline}
where ${\ve{u}} = \lbrace t, \ve{r}, {\ve{p}} \rbrace$ denotes extended phase-space coordinates and $f({\ve{u}})$ is the (spin-independent) distribution function. In the following, we are interested in the homogeneous, stationary solution, such that $f({\ve{u}})\equiv f({\ve{p}})$. The left-hand side of Eq.~\eqref{eq:BE_full} includes the effect of temperature gradients and fields through the substantial time derivative. The right-hand side includes in our case the input from impurity and electron-electron scattering. Assuming Matthiesen's rule to apply, we can treat the two contributions separately. Impurity scattering is formulated as
\begin{multline}\label{eq:CI}
	\big[\partial_t f({\ve{p}})\big]_{\rm imp} = - \Omega\int (d{\ve{p}}') \Gamma^{\rm imp}_{{\ve{p}}, {\ve{p}'}}\times\\ 
		\big\{f({\ve{p}})\big[1 - f({\ve{p}'})\big]-\big[1 - f({\ve{p}})\big]f({\ve{p}'})\big\},
\end{multline}
where $(d{\ve{p}}) = d{\ve{p}} / (2\pi\hbar)^2$, $\Omega$ is the sample volume and scattering rates are determined by the Fermi Golden rule,
\begin{equation}\label{eq:CI_Gamma_Fermi}
  \Gamma^{\rm imp}_{{\ve{p}},{\ve{p}'}}= \frac{2\pi}{\hbar}n_{\rm imp}{|v_{\rm imp}|}^2
\delta(\varepsilon_{\ve{p}}-\varepsilon_{\ve{p}'}).
\end{equation}
Here, $n_{\rm imp}$ is the density of impurities and $v_{\rm imp}=\langle {\ve{p}}|\hat{V}_{\rm imp}|{\ve{p}'}\rangle$ is the scattering matrix element, which we assume to be isotropic, considering only $s$-wave (contact) scattering. We introduce the dimensionless impurity scattering strength $n_{\rm imp}|v_{\rm imp}|^2/t^2=0.01$, where $t$ is the characteristic energy scale of the electron dispersion $\varepsilon_{\ve{p}}$ as described in  App.~\ref{app:params}.

The effect of collision between electrons is included in the second term on the right hand side of Eq.~\eqref{eq:BE_full}, 
\begin{align}\label{eq:CI_el_el}
	\big[\partial_t f({{\ve{p}}_1})\big]_{el-el} =&- \Omega^3\int(d{{\ve{p}}_2})(d{{\ve{p}}_3})(d{{\ve{p}}_4}) \Gamma^{\rm el-el}_{{\ve{p}}_1,{\ve{p}}_2,{\ve{p}}_3,{\ve{p}}_4}\times\nonumber\\
	&\Big\{ f({\ve{p}}_1)f({\ve{p}}_2)\big[1-f({\ve{p}}_3)\big]\big[1-f({\ve{p}}_4)\big]\nonumber\\
	&-\big[1-f({\ve{p}}_1)\big]\big[1-f({\ve{p}}_2)\big]f({\ve{p}}_3)f({\ve{p}}_4)\Big\}
\end{align}
whereby the two electrons change their momenta, $ (\ve{p}_1 , \ve{p}_2 ) \leftrightarrow (\ve{p}_3 , \ve{p}_4 ) $.  Scattering rates are computed via the Fermi Golden rule taking a repulsive on-site Hubbard-U-type coupling\cite{buhmann:2013b} (we fix $U = 2t$), which yields
\begin{align}\label{eq:Gamma_el_el}
 \Gamma^{\rm el-el}_{{\ve{p}}_1,{\ve{p}}_2,{\ve{p}}_3,{\ve{p}}_4} \propto & U^2 \delta(\varepsilon_{{\ve{p}}_1}+\varepsilon_{{\ve{p}}_2}-				\varepsilon_{{\ve{p}}_3}-\varepsilon_{{\ve{p}}_4})\nonumber\\
 &\times\delta({\ve{p}}_1+{\ve{p}}_2-{\ve{p}}_3-{\ve{p}}_4)
 \end{align}
and satisfies energy as well as momentum conservation. While the scattering rate is isotropic ($s$-wave scattering) we find a highly anisotropic contribution to Eq.~\eqref{eq:CI_el_el} because of Umklapp scattering--the fact that momentum conservation allows for momentum transfer of reciprocal lattice vectors. Indeed, Umklapp scattering is the only way of momentum relaxation for the electron-electron collision.  

The final ingredient for our calculations is the electron dispersion, which we model after the $\gamma$ band of Sr$_2$RuO$_4$ within a tight-binding description, see App.~\ref{app:params}. In the dispersion, we include the effect of doping and uniaxial strain through the chemical potential and the hopping integrals, respectively. Note that we use a non-trivial Poisson ratio in our calculations, which leads to an asymmetric response to positive and negative uniaxial strains.

To simplify Eq.~\eqref{eq:BE_full}, we use the parametrization~\cite{buhmann:2013c}
\begin{equation}
 	f({\ve{p}}) = \left[1 + \exp{\left(\frac{\varepsilon({\ve{p}}) - \mu}{T} - \phi({\ve{p}})\right)}\right]^{-1},
\end{equation}
which yields in linearized form
\begin{equation}\label{eq:f_1order}
	f({\ve{p}}) \approx f_{0}({\ve{p}})  + \underbrace{f_{0}({\ve{p}})\left[1 - f_{0}({\ve{p}})\right]}_{\textrm{scattering phase space}}\phi({\ve{p}}),
\end{equation}
with $ f_{0}({\ve{p}}) $ the Fermi-Dirac distribution. The correction $\phi({\ve{p}})$ to be determined, thus, directly relates to the scattering phase space. This correction contains the necessary information to calculate transport coefficients such as the longitudinal electronic conductivity for an electric field $ E $ along the $x$ axis,
\begin{equation}
	\sigma_{xx} = \frac{e}{E} \int (d{\ve{p}})f_0({\ve{p}})\big[1-f_0({\ve{p}})\big]\phi({\ve{p}})v_x({\ve{p}})
	\label{eq:el_cond}
\end{equation}
and the longitudinal thermal conductivity for a temperature gradient $ T' = ( \ve{\nabla} T)_x $ along the $x$ axis,
\begin{equation}
	\kappa_{xx} = \frac{1}{T'} \int (d{\ve{p}}) f_0({\ve{p}})\big[1-f_0({\ve{p}})\big]\phi({\ve{p}})v_{x}({\ve{p}}) (\varepsilon_{\ve{p}}-\mu).
	\label{eq:th_cond}
\end{equation}
Before solving the full Boltzmann equation including electron-electron scattering numerically in Sec.~\ref{sec:numerics}, we first analyze the low-temperature limit where impurity scattering dominates.

\section{Impurity scattering}
\label{sec:impurities}

Focusing on impurity scattering only we obtain the linearized Boltzmann equation for $ \phi({\ve{p}}) $ up to linear order in a constant electric field ${\ve{E}}$ and a temperature gradients ${\ve{\nabla}}_{\ve{r}}T$. As shown in App.~\ref{app:impurities}, the resulting corrections reads 
\begin{equation}\label{eq:lin_stat}
	\phi({\ve{p}})=-\left[\left(\frac{\varepsilon_{\ve{p}}-\mu}{T}\right){\ve{\nabla}}_{\ve{r}}T + e{{\ve{E}}}\right]\cdot\frac{{\ve{v}}_{\ve{p}}}{T}\tau(\varepsilon_{\ve{p}}).
\end{equation}
Here, ${\ve{v}}_{\ve{p}} \equiv \dot{\ve{r}}= \partial_{\ve{p}}\varepsilon_{\ve{p}}$ is the velocity and the effect of impurity scattering appears in the scattering time $\tau(\varepsilon) = \hbar/\big[2\pi n_{\rm imp} |v_{\rm imp}|^2 \Omega N(\varepsilon)\big]$ with $N(\varepsilon)$ the density of states at energy $\varepsilon$. Note that for $s$-wave scatterers the scattering time is not direction dependent.

Sufficiently far from the vHS, the density of states $N(\varepsilon)\approx N$ is only weakly depending on energy, resulting in an essentially constant scattering time $\tau$. Therefore, we recover the well-known scattering-time approximation~\cite{ziman:1972}
\begin{eqnarray}\label{eq:limit_cases}
	f_{{\ve{E}}}({\ve{p}}) \approx f_0\left({\ve{p}} + \frac{e\tau}{\hbar}{{\ve{E}}}\right), \\
	f_{{\ve{\nabla}} T}({\ve{p}})\approx \frac{1}{1+e^{\frac{\varepsilon_{\ve{p}}-\mu}{T-\tau{\ve{v}}\cdot {\ve{\nabla}} T}}}. 
\end{eqnarray}

As we approach the vHS, we find that $\phi({\ve{p}})$ vanishes for all directions of $\ve{p}$, since $\tau(\varepsilon)\propto 1/N(\varepsilon) \to 0 $. Thus, we expect for both types of Lifshitz transitions a qualitatively similar behavior, namely a suppression in all directions due to the vanishing scattering time. Physically, the large scattering phase space available at energies near the vHS leads to a fast momentum relaxation such that an applied field or temperature gradient can shift the Fermi distribution only weakly. Note that this effect is naturally weaker in the case of a Lifshitz transition at only one van Hove point as realized by uniaxial strain.

\begin{figure}[tpb]
\includegraphics[width=8.7 cm]{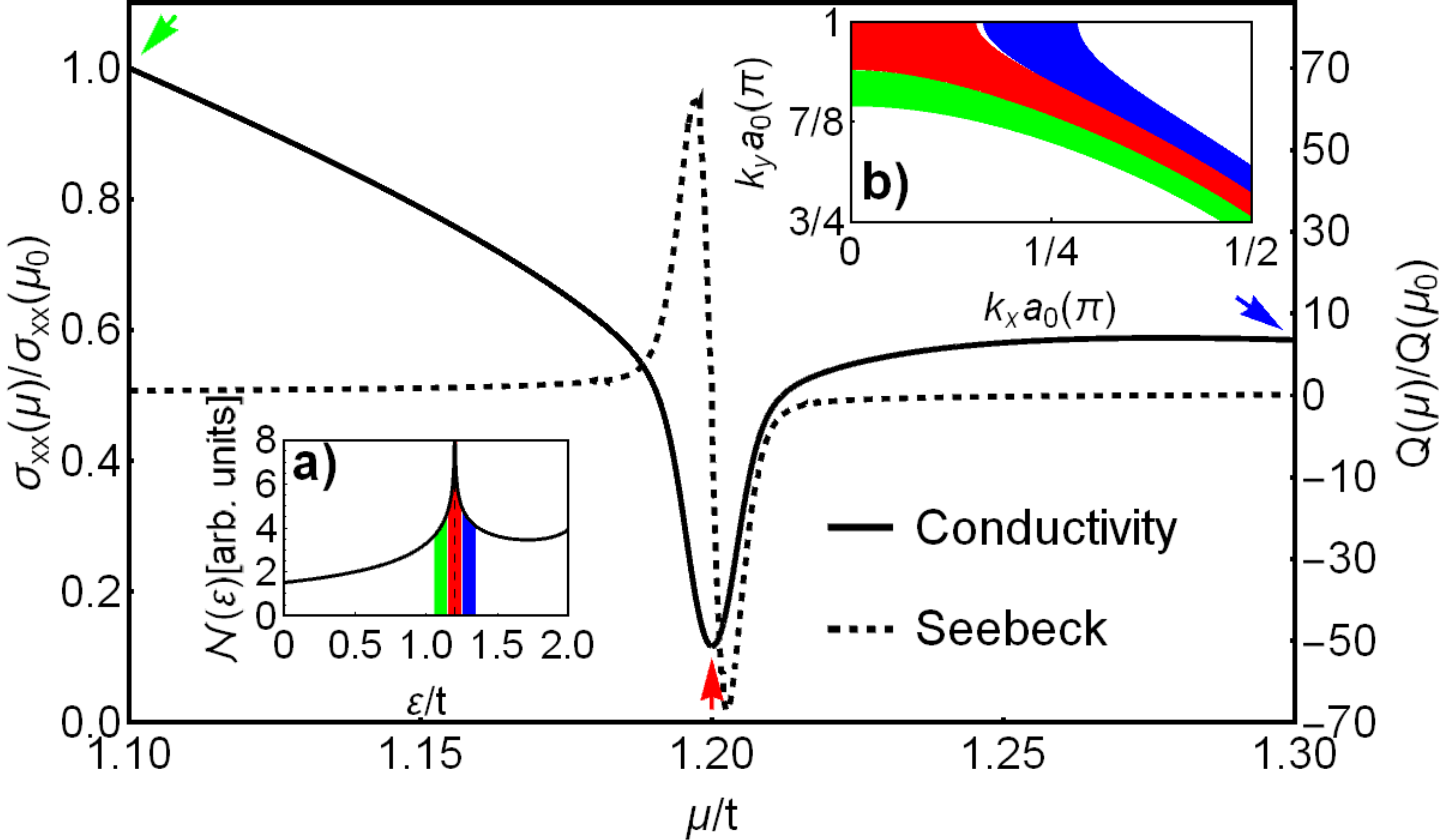}
\caption{
Conductivity dip due to enhanced impurity scattering together with the switching sign of the Seebeck coefficient near the Lifshitz point normalized to the undoped resistivity $\mu_0 = 1.1 t$. Inset a) shows the density of states together with the energy intervals of the contributing states for $\mu/t = 1.1$ (green), $\mu/t = 1.2$ (red), $\mu/t = 1.3$ (blue). Inset b) shows the contributing states for these three cases for the scattering phase space $f_0({\bf p})\big(1-f_0({\bf p})\big)\leq 10^{-10}$.
}
\label{Fig:Dip_Simple}
\end{figure}

For a qualitative understanding of electrical transport at low temperatures and Fermi energy near a vHS, we use Eq.~\eqref{eq:lin_stat} to obtain the conductivity through Eq.~\eqref{eq:el_cond},
\begin{eqnarray}
	\sigma_{xx} &=& \frac{e^2}{T} \int (d{\ve{p}})f_0({\ve{p}})\big[1-f_0({\ve{p}})\big]\tau(\varepsilon_{\ve{p}})\big(v^x_{\ve{p}}\big)^2\label{eq:imp_cond_0}\\
	&=& \frac{e^2}{T} \int d\varepsilon f_0(\varepsilon)\big[1-f_0(\varepsilon)\big] \tau(\varepsilon)\Big\langle \frac{\big(v^x_{\ve{p}}\big)^2}{|{\ve{\nabla}}_{\ve{p}} \varepsilon_{\ve{p}}|}\Big\rangle_\varepsilon,\label{eq:imp_cond}
\end{eqnarray}
where we have introduced the equal-energy-contour average $\langle A \rangle_\varepsilon = \int_{\varepsilon_{\ve{p}}=\varepsilon}(d{\ve{p}}) A$. 
For $T\rightarrow 0$, this integral is dominated by $ \varepsilon = \varepsilon_F $, the Fermi energy, such that
\begin{equation}
	\sigma_{xx} \approx e^2 \tau(\varepsilon_{\rm F})\Big\langle \frac{\big(v^x_{\ve{p}}\big)^2}{|\ve{v}_{\ve{p}}|}\Big\rangle_{\varepsilon_{\rm F}},
	\label{eq:sigma_xx_0}
\end{equation}
where $\langle\cdot\rangle_{\varepsilon_{\rm F}}$ denotes a Fermi surface average. As noted above, the suppression of the scattering time $  \tau(\varepsilon_{\rm F}) $ close to the vHS leads to a reduction of the conductivity.
Figure~\ref{Fig:Dip_Simple} shows the corresponding dip in the longitudinal conductivity by evaluating Eq.~\eqref{eq:imp_cond_0} together with the Seebeck coefficient approximated through Mott's formula
\begin{equation}
Q = -\left.\frac{\pi^2}{3}\frac{k_B^2 T}{e}\frac{\sigma'(\varepsilon)}{\sigma(\varepsilon)} \right|_{\varepsilon = \mu},
\label{eq:Mott-1}
\end{equation}
considering the doping scenario at a low temperatures $(T/t\sim 10^{-3})$. Note that the shown conductivity is already in good qualitative agreement with the experimental findings of Refs.~\onlinecite{barber:2018} and \onlinecite{burganov:2016}.

An interesting feature arises for the case where uniaxial strain pushes the Fermi surface through van Hove points at $\ve{k} = (0,\pi)$ corresponding to negative $\varepsilon_{xx}$ in Fig.~\ref{Fig:DOS_Gamma_Band_Umklapp} b). In this case, the averages over the energy contours in Eq.~\eqref{eq:imp_cond} are dominated by the momentum direction along $ [1,0] $. 
For $T=0$, the conductivity vanishes for $ \varepsilon_{xx} = \varepsilon_{\rm vH}$ according to Eq.(\ref{eq:sigma_xx_0}) due to the vanishing scattering time. Once the temperature increases, the range of energy averaging grows such that with $ \tau $ also the conductivity becomes finite and grows with increasing temperature. While upon uniaxial deformation ($ \varepsilon_{xx} <0 $) the Fermi surface along $[0,1]$ passes through the vHS, it retreats from the van Hove point along $ [1,0]$ due to Luttinger's theorem. Therefore, the Fermi velocity $ v_{\rm F}^x $ along $ [1,0]$ increases monotonically with increasing deformation. Since the scattering time $ \tau $ is an approximately symmetric function of $ \varepsilon_{xx} - \varepsilon_{\rm vH}$ with its minimum at $ \varepsilon_{xx} = \varepsilon_{\rm vH}$, the $ \varepsilon_{xx} $-dependence of the average of $ \langle \big(v^x_{\ve{p}}\big)^2/|\ve{v}_{\ve{p}}| \rangle_{\varepsilon} $ leads to a shift of the minimum towards $ \varepsilon_{xx} < \varepsilon_{\rm vH}$ for increasing temperature. This behavior agrees well with the numerical solution of the linearized Boltzmann equations presented in the next section and reproduces also the qualitative behavior observed in experiments \cite{barber:2018}. For the opposite situation of positive uniaxial strain, $ \varepsilon_{xx} > 0 $, this argument does not apply and we will see below that there is no temperature-dependent shift of the conductance minimum.

\section{Numerical results}
\label{sec:numerics}
\begin{figure}[htpb]
\includegraphics[width = 7 cm]{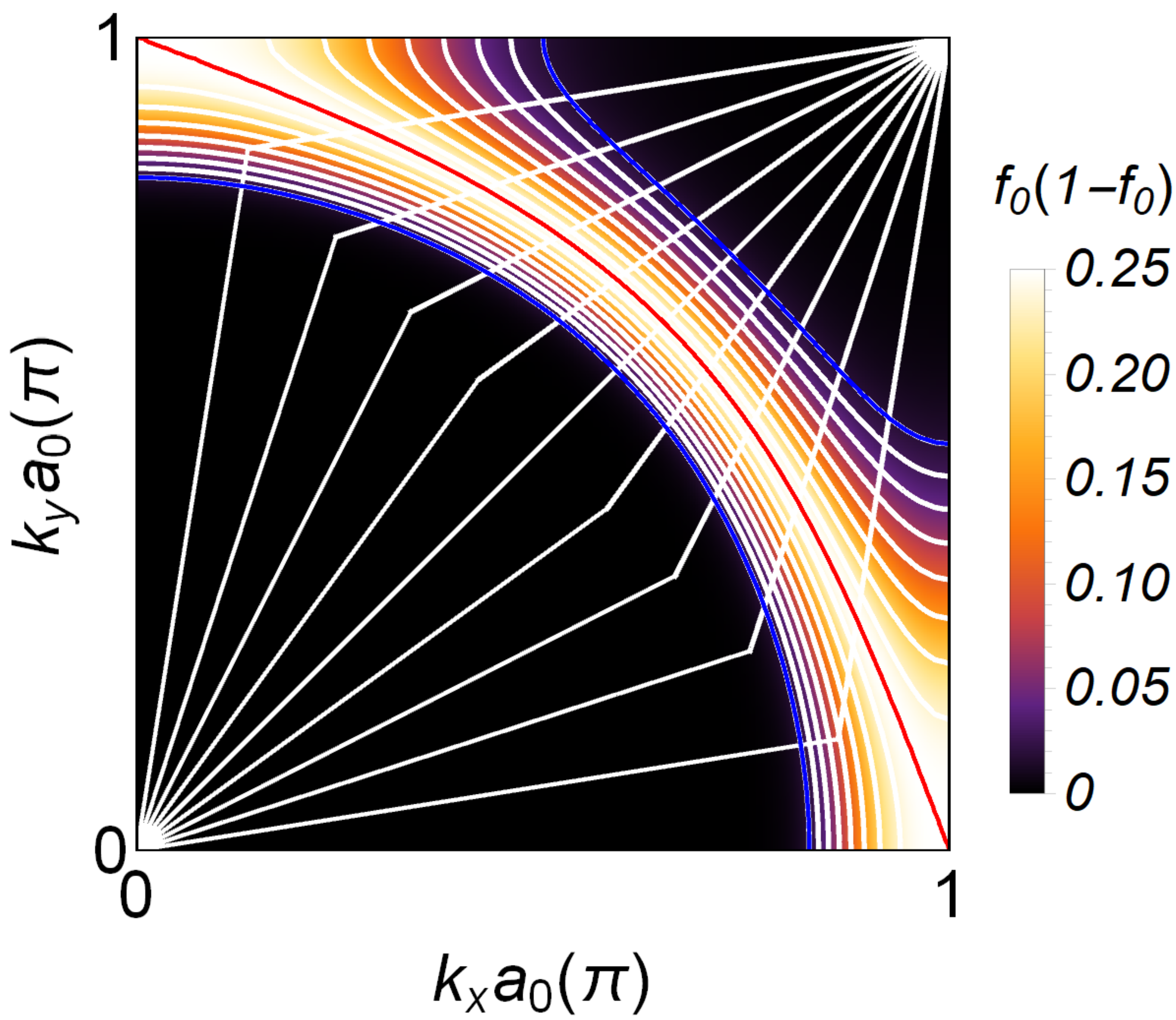}
\caption{
Discretization of the first BZ adapted to the scattering phase space. The red curve corresponds to the Fermi surface crossing the VHS. Blue curves denote the boundary equipotentials of the discretization, i.e., $\varepsilon_{\rm F}-4T$ and $\varepsilon_{\rm F}+4T$. Note that for demonstration purposes, a large temperature $\sim 100K$ is shown.}
\label{Fig_discret}
\end{figure}

We now turn to the numerical solution of the linearized Boltzmann Eq.~\eqref{eq:BE_full} including both impurity and electron-electron scattering. The most important part of momentum space for this calculation is close to the Fermi surface representing the scattering phase space. Thus, we adopt the discretization scheme from Ref.~\onlinecite{buhmann:2013a} shown in Fig.~\ref{Fig_discret}, with momentum space patches following band energy equipotential lines distributed between $\varepsilon_F - 4T$ and $\varepsilon_F + 4T$ together with equally distributed angular $\vartheta$ coordinates. 
An advantages of our technique is that the patched discretization of the BZ is adaptive to the temperature. In other words, we always work with the same number of patches regardless of the temperature. Note that we normalize the temperature scale with respect to $t = 0.14$ eV~\cite{hsu:2016} in order to compare with experimental results. 

\begin{figure*}[htpb]
\includegraphics[width=16 cm]{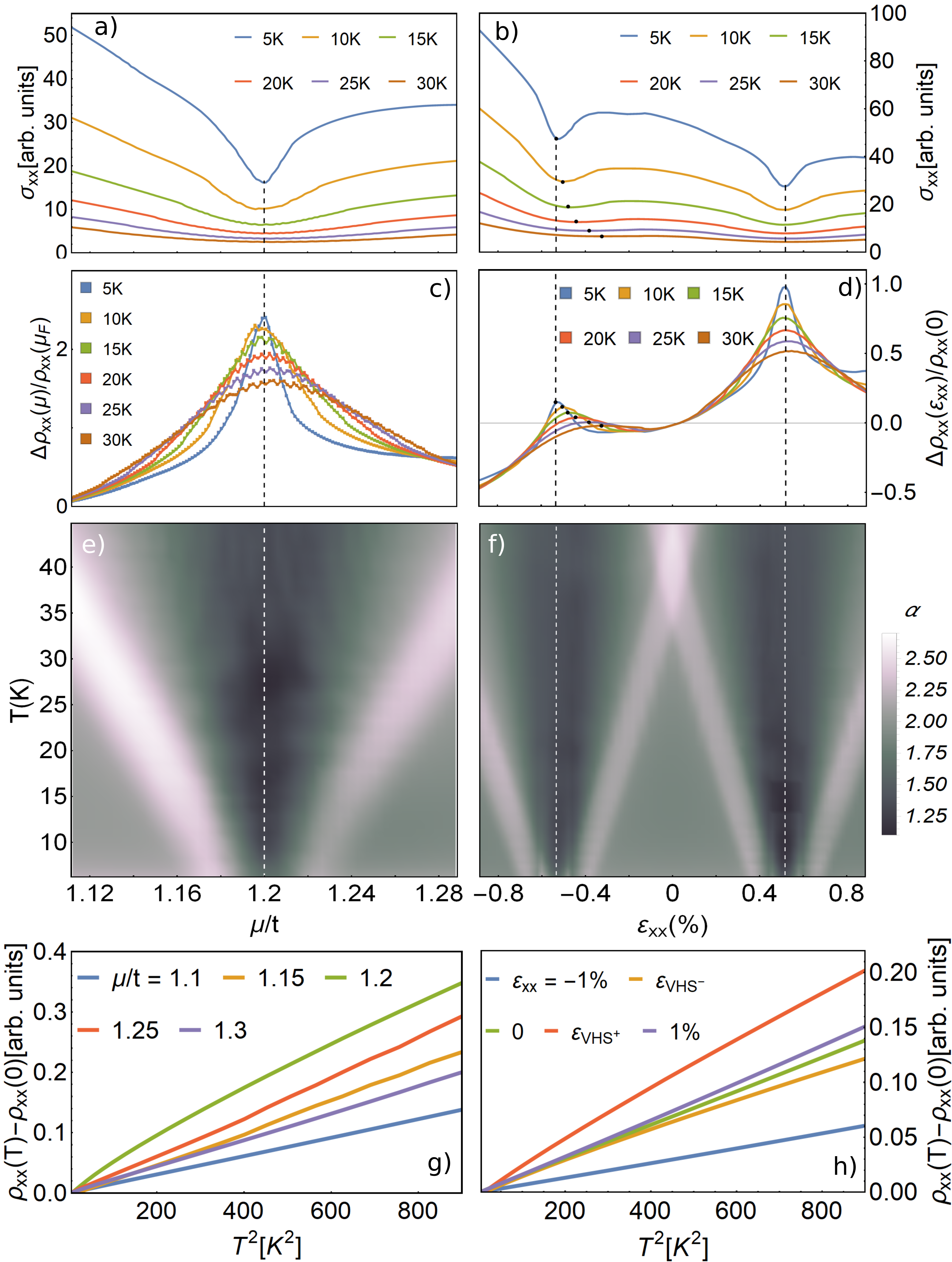}
\caption{
Summary of numerical results for the electronic transport:
Temperature evolution of the conductivity [a), b)], normalized resistivity [c), d)], Temperature scaling coefficient $\alpha$ [e), f)],  as well as resistivity curves for various values of $\mu/t$ and $\varepsilon_{xx}$ comparing band filling (left panel) and uniaxial strain (right panel) scenarios. Subfigures g) and h) correspond to temperature cuts of the resistivity at values of interesting band fillings and values of uniaxial strains. Notice the strongly nonlinear temperature dependence of $\rho_{xx}(T^2)$ considering $\mu/t = 1.2$ [green curve in subfigure g)] and similar dependence at the both Lifshitz points $\varepsilon_{vHS^{\pm}}$ [orange and red curves in subfigure h)].
}
\label{Fig:BF&US_electrical}
\end{figure*}

We use a sufficiently dense set of angular ($\vartheta$) and  energy ($\varepsilon$) contours, namely 160 angular and 30 energy contours, to ensure high accuracy. A high-resolution discretization is particularly important for the calculation of the electron-electron collision integral, Eq.~\eqref{eq:CI_el_el}, and its anisotropy due to Umklapp processes~\cite{buhmann:2013b}, where the relevant phase space is located around the crossing points of the Fermi surfaces and the Umklapp zones denoted in Fig.~\ref{Fig:DOS_Gamma_Band_Umklapp}. We distinguish two Umklapp processes, one for reciprocal lattice vectors $ (2 \pi,0) $ and $ (0,2 \pi) $ corresponding to the boundaries of the dark blue zones in Fig.~\ref{Fig:DOS_Gamma_Band_Umklapp} and the $ (2\pi,2\pi) $ indicated by the white diamond-shaped boundary. Only such high resolution allows us to analyze the subtle low-temperature dependence of transport quantities which are strongly influenced by the position of the Fermi surface crossings. 
Only when we generate maps (and low temperature details) of quantities such as shown in Fig.\ref{Fig:BF&US_electrical} e), f) do we reduce the angular resolution to 40 contours for performance reasons. This resolution is sufficient in these cases to provide information on temperature dependence of the displayed quantities.   

\subsection{Electrical transport}
Figure~\ref{Fig:BF&US_electrical} provides an overview of our main numerical results for the electrical conductivity, $ \sigma_{xx} $, for the two cases of Fermi surface tuning using the full solution to Eq.~\eqref{eq:BE_full}. The upper six panels provide scans of the tuning parameter $ \mu $ for the case of doping on the left-hand side and $ \varepsilon_{xx} $ for uniaxial deformation on the right-hand side. The Lifshitz transitions occur at $ \mu = 1.2 t $ and $ \varepsilon_{xx} = \varepsilon_{\rm vH} \sim \pm 0.53 \% $. Panels a) and b) display the conductivity with dips, which become more pronounced with decreasing temperature, at the Lifshitz transitions. The shift of the minimum away from the Lifshitz-transition point in the case $ \varepsilon_{xx}<0 $ is marked by dots in panel b). Panels c) and d) show the relative change of the resistivity $ \Delta \rho_{xx} (\mu) = \rho_{xx} (\mu) -\rho_{xx}(\mu_F) $ normalized with respect to $ \rho_{xx}(\mu_F) $, where $ \mu_F = 1.1 t $ is the chemical potential of the undoped case, and $ \Delta \rho_{xx} (\varepsilon) = \rho_{xx} (\varepsilon) -\rho_{xx}(0) $ normalized with respect to $ \rho_{xx}(\varepsilon_{xx} = 0) $. In both cases, the dips in $ \sigma_{xx} $ translate to peaks whose maxima grow with decreasing temperature. Note that for $ \varepsilon_{xx} < 0 $ the resistivity, including the shift of the maxima, resembles qualitatively very well the experimental results found in Ref.~\onlinecite{barber:2018}. 

We turn to the low-temperature behavior of $ \rho_{xx} $, which we use as our first tool to identify deviations from the standard Fermi-liquid picture. The temperature dependence of the resistivity, as displayed in the  panels g) and h) can be fitted by
\begin{equation}
\rho_{xx} (T) = \rho_{0}+A T^{\alpha},
\label{ATalpha}
\end{equation}
with the parameters $A$ and $ \alpha $ besides the residual resistivity $ \rho_0 $. The exponent $ \alpha = 2 $ denotes a Fermi liquid, while $ \alpha $ smaller than 2 is considered as non-Fermi liquid behavior. We determine the exponent from the numerical results using
\begin{equation}
	\alpha = \frac{\partial \ln\big[\rho(T)-\rho(0)\big]}{\partial \ln T},
	\label{alpha}
\end{equation} 
which yields the $ \alpha $-maps in panels e) and f). In both cases, we see triangle-shaped regions with $ \alpha \approx 2 $ far enough from the Lifshitz points. On the other hand, at the Lifshitz transition a fan of values $ \alpha $ clearly smaller than 2 opens. The two regimes are separated by bright stripes ($ \alpha > 2 $), which result from Eq.~(\ref{alpha}) when  $ \rho_{xx} $ has a kink-like feature as the vHS enters the scattering phase space upon increasing the temperature. These kinks are visible by eye in the plots of $ \rho_{xx} $ in panel g) for $ \mu/t = 1.15 $ and $ 1.25 $.

\begin{figure*}[htpb]
\includegraphics[width = 16 cm]{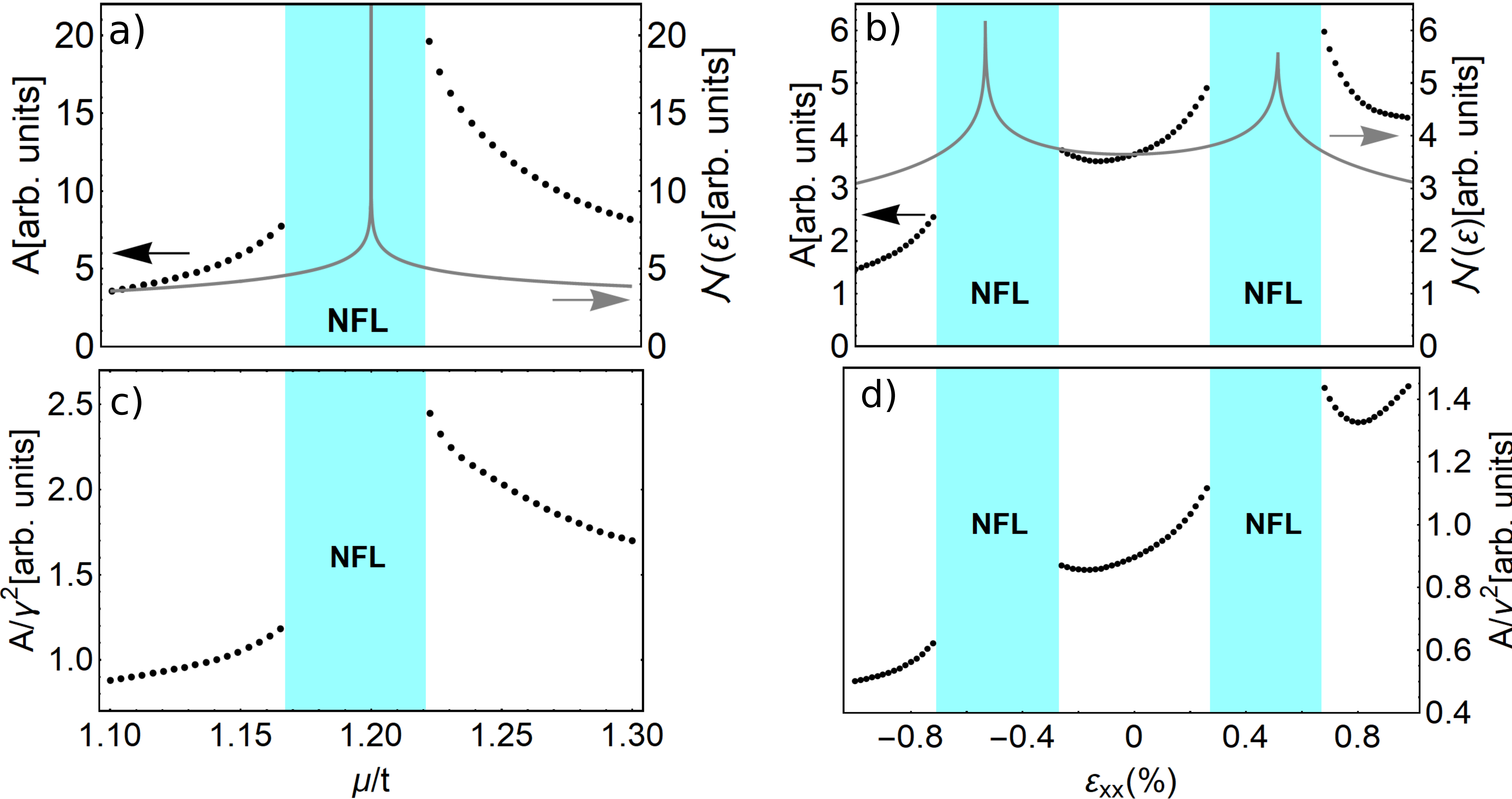}
\caption{
Coefficient A of $\rho(T) = \rho_0+ AT^2$ [subfigures a) and b)] together with the Kadowaki-Woods ratio $R_{\rm KW}= A/\gamma^2$[subfigures c) and d)] with as a function of band filling (left panel) and uniaxial strain (right panel). Colored regions correspond to the non-Fermi liquid behavior of the resistivity ($\alpha\neq 2$), outside of which the density of states (gray lines) remains almost constant.}
\label{Fig:BF&US_KW}
\end{figure*}

\subsection{Kadowaki-Woods ratio}
In the Fermi-liquid regime ($ \alpha = 2 $), the fitting parameter $A$ in Eq.~\eqref{ATalpha} can be used together with the Sommerfeld coefficient $\gamma\propto N(\varepsilon_{\rm F})$ to define the Kadowaki-Woods ratio $R_{\rm KW}= A/\gamma^2$. This ratio is empirically a material-independent constant for several material classes~\cite{kadowaki:1986} and has been suggested to remain constant, if the strength of the
electron correlations varies for a fixed bare band structures~\cite{jacko:2009}. 
In their experiments with uniaxially deformed Sr$_2$RuO$_4$, Barber et al. observed an increase of $A$ faster than that of $ \gamma^2$, anticipated from DFT calculations, upon tuning the Fermi surface towards the vHS for $ \varepsilon_{xx} < 0 $.  In order to analyze this behavior, we determine $ A $ and $ R_{\rm KW} $ within our approach, where we calculate $ \gamma $ from our tight-binding band structure. 
Figure~\ref{Fig:BF&US_KW} shows both $ A $ and $ R_{\rm KW} $ in the low-temperature regime, where $ \alpha = 2 $ indicates Fermi-liquid-like behavior. We find that $ A $ increases as the Fermi surface shifts towards the vHS and indeed faster than $ \gamma^2 $ such that the Kadowaki-Woods ratio also increases in a similar way. Note that the behavior we find for $ A $ considering $ \varepsilon_{xx} < 0 $ in panel b) is in good qualitative agreement with the experimental findings\cite{barber:2018}.

\subsection{Thermal transport \& Seebeck coefficient}

\begin{figure*}[htpb]
\includegraphics[width = 16 cm]{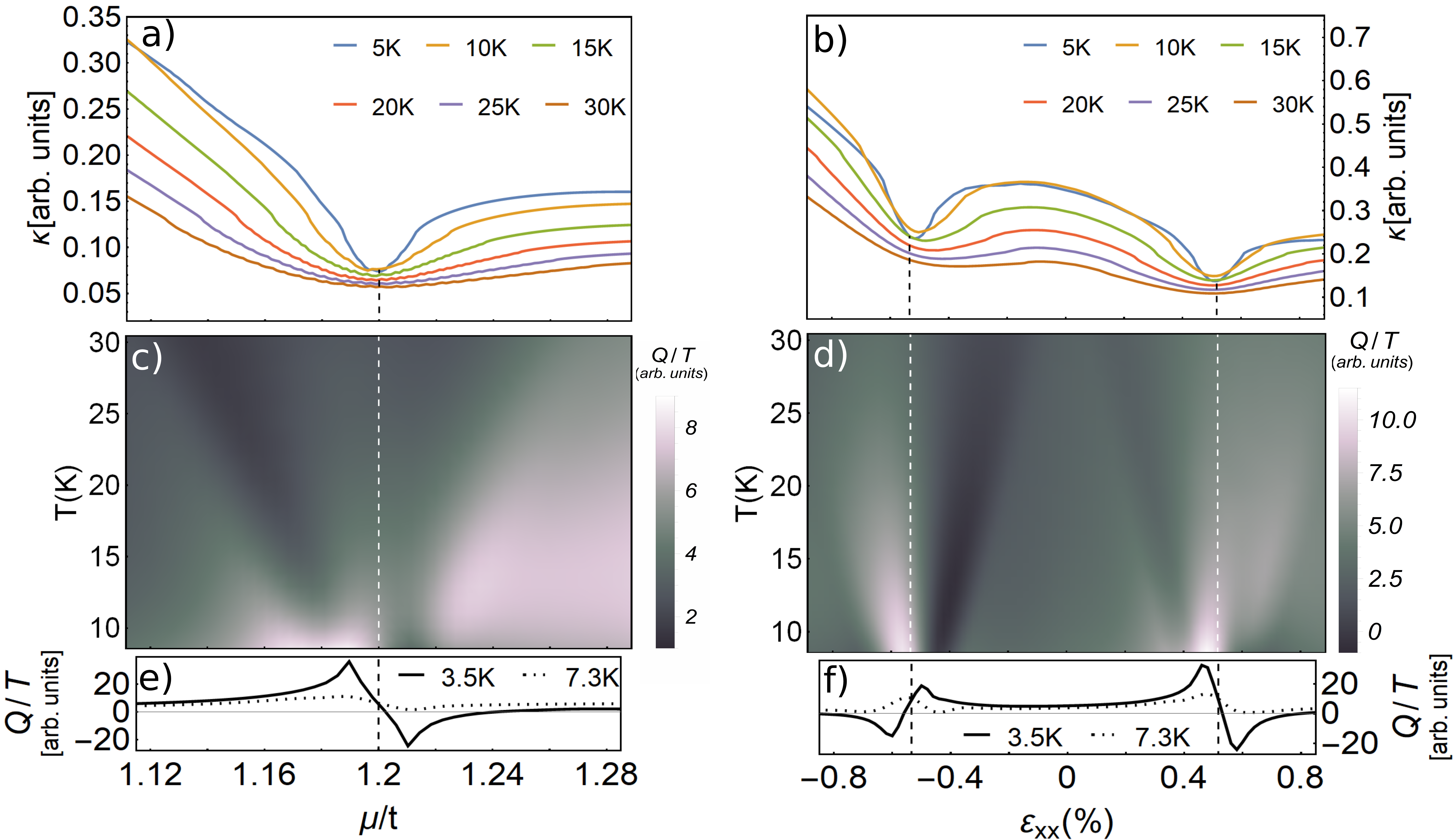}
\caption{Temperature evolution of the thermal conductivity [a) and b)], as well as Seebeck coefficient divided by the temperature $Q/T$ [c) and d)] together with the detail at the low temperature $Q/T$ [e) and f)] comparing the band-filling and the uniaxial-strain scenario.}
\label{Fig:BF&US_thermal_and_so}
\end{figure*}

The electronic contribution to the thermal transport can be calculated numerically in a manner analogues to the electrical conductivity in the previous section. Figures~\ref{Fig:BF&US_thermal_and_so}  a) and b) show the thermal conductivity $ \kappa_{xx} $, which exhibits similar features as the conductivity $ \sigma_{xx} $ presented in Fig.~\ref{Fig:BF&US_electrical}, including, in particular, the dips at the Lifshitz transitions. These dips are naturally wider, because the integral for the thermal conductivity
\begin{equation}
	\kappa_{xx} = \frac{1}{T'} \int (d{\ve{p}}) f({\ve{p}}) v^{x}_{\ve{p}} \big(\varepsilon({\ve{p}})-\mu\big),
\end{equation}
contains also the electron dispersion which is essentially flat at the vHS. A further difference is the weaker temperature dependence of the minima. However, there is again a temperature dependent shift of the minimum position for uniaxial strain $ \varepsilon_{xx} < 0 $ as observed and discussed for $ \sigma_{xx} $ [see Fig.~\ref{Fig:BF&US_electrical} b)]. 

The Seebeck coefficient $Q$ relates a temperature gradient to a resulting electric field, $\ve{E} = Q {{\ve{\nabla}}} T$ under open circuit condition and follows from the solution to the linearized Boltzmann equation. It is given by
\begin{equation}
Q = \frac{| \ve{E}|}{e | \ve{\nabla} T|} \frac{\int (d{\ve{p}}) v^x_{\ve{p}}f_0(\ve{p})\big(1-f_0(\ve{p})\big)\tilde{\phi}_{\ve{\nabla T}}(\ve{p}) }{\int (d{\ve{p}}) v^x_{\ve{p}}f_0(\ve{p})\big(1-f_0(\ve{p})\big)\tilde{\phi}_{\ve{E}}(\ve{p})}, 
\end{equation}
where $\tilde{\phi}_{\ve{E}}$ and $ \tilde{\phi}_{\ve{\nabla T}} $ correspond to the solution of the Boltzmann equation with only external electric field $\ve{E}$ or temperature gradient $\ve{\nabla T}$. 
We can use the Seebeck coefficient as a second tool to identify deviations from Fermi-liquid behavior, since for a Fermi liquid, $ Q $ has linear temperature dependence. For this reason, we plot the ratio $ Q/T $ in 
Figs.~\ref{Fig:BF&US_thermal_and_so} c) and d) as a map of temperature versus chemical potential and strain. We again recognize different regimes, where the Umklapp processes are clearly observable in both panels. In the panels e) and f), we focus on the low temperature regime and show $ Q/T $ scans for fixed temperatures. We find little temperature dependence away from the Lifshitz points in accordance with expectations for Fermi liquids. Around the Lifshitz points, on the other hand, anomalies emerge with lowering temperature including a sign change. Interestingly, this agrees well with the Mott formula given by Eq.~\eqref{eq:Mott-1} considering the case of tuning by chemical potential. The dip in conductance results in a sign change with the anomalies corresponding to the inflection points of $ \sigma_{xx}(\mu) $. The reason for this agreement lies in the disappearance of the highly anisotropic contribution of electron-electron scattering to momentum relaxation at low temperature. At higher temperature these anisotropies change the behavior of $ Q/T $ rather profoundly \cite{buhmann:2013c,buhmann:2013d}. The same argument applies also to the case of Fermi-surface tuning by uniaxial strain, where, near the Lifshitz point, the Mott formula can be approximated by
\begin{equation}
Q \propto - \frac{1}{\sigma(\varepsilon_{xx})} \frac{\partial \sigma(\varepsilon_{xx})}{\partial |\varepsilon_{xx}|} .
\end{equation}
Thus, a change of $ \varepsilon_{xx} $ has an analogous effect as a variation of the chemical potential, justifying this approximation. 

\begin{figure*}[htpb]
	\includegraphics[width = 16 cm]{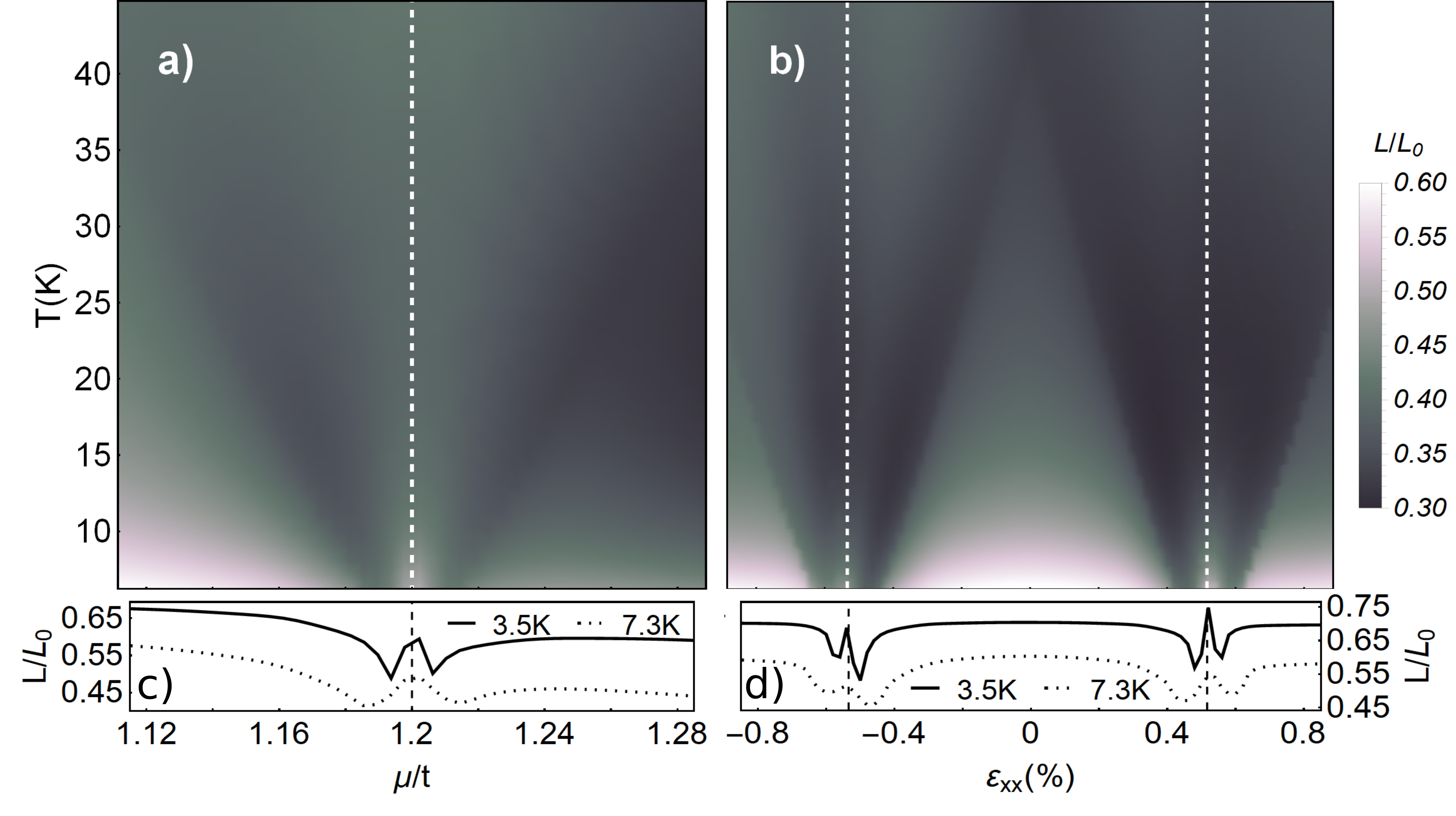}
	\caption{
		Deviations from the Wiedemann-Franz law as a function of band filling (left panel) and uniaxial strain (right panel). The 					dashed lines mark the Lifshitz transitions.
	}
	\label{Fig:BF&US_WF}
\end{figure*}

\begin{figure*}[htpb]
\includegraphics[width = 16 cm]{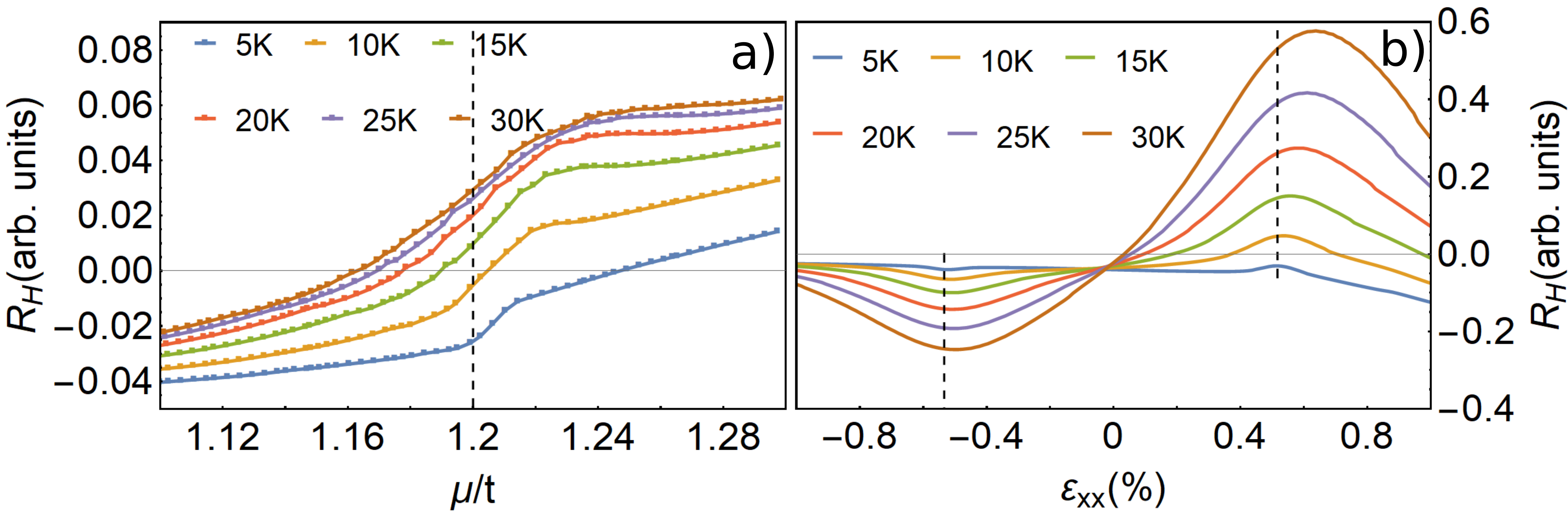}
\caption{
Temperature dependence of the Hall resistivity for the band-filling as well as the uniaxial-strain scenario. The line at $R_{\rm H}=0$ denotes where the character of the charge carriers changes.
}
\label{Fig:BF_R_H}
\end{figure*}

\subsection{Wiedemann-Franz law}

The anisotropy of scattering due to Umklapp processes affects the electrical and thermal conductance in a different way. This impacts the temperature dependence of both, as well as the Wiedemann-Franz law. The Lorenz, number defined as
\begin{equation}
L(T) = \frac{\kappa_{xx}(T)}{T \sigma_{xx}(T)},
\end{equation}
is constant and given by $ L_0 = \pi^2 k_B^2/3e^2 $ for isotropic scattering. In Fig.~\ref{Fig:BF&US_WF}, we display the ratio $ L(T)/L_0 $ for a range of varying band fillings [a)] and uniaxial deformation [b)] including the Lifshitz transitions.  A general observation is that $ L < L_0 $, signaling that the thermal transport is more efficiently impeded by scattering here than the charge transport. In the regions far from the Lifshitz transitions, which we identified previously as the (triangle-shaped) Fermi liquid regime, $ L/L_0 $ is larger and additionally increases as temperature decreases, such that $ L $ approaches $ L_0 $.  This indicates that electron-electron scattering is responsible for the suppression of $ L $, while isotropic impurity scattering leads to a higher value of $ L $. Notable is the difference between the Fermi liquid regions for $ \mu < 1.2t $ and $ \mu > 1.2 t $ in Fig.~\ref{Fig:BF&US_WF}a). While in the former case both Umklapp processes, involving the reciprocal lattice vectors of the kind of $ (2 \pi,0 ) $ and $ (2 \pi , 2 \pi) $, are allowed, in the latter case only the $ (2 \pi , 0 )$-type of Umklapp contributes and leads to a more pronounced anisotropy. At the same time, the electron density, which increases with the chemical potential, increases the scattering probability. This leads to a stronger suppression of $ L $ than in the range $ \mu < 1.2t $. In contrast, the Fermi liquid regimes for a uniaxial deformation are fairly comparable in their behavior and the whole picture looks rather symmetric around $ \varepsilon_{xx} $. 

In the vicinity of the Lifshitz points, we observe a pronounced drop of $ L/L_0 $, whose onset is temperature dependent and can be identified by the overlap of the scattering phase space with the vHS. This enlarged phase space is accompanied by more electronic states contributing to energy transport. This is supported by the observation that, comparing Figs.~\ref{Fig:BF&US_electrical} a) and b) with Figs.~\ref{Fig:BF&US_thermal_and_so} a) and b), the dips in the conductivity are sharper in the former than the latter case. This gives rise to the strong reduction of $ L $. Right at the Lifshitz point, on the other hand, we find a very narrow region where $ L $ recovers for very low temperatures. This effect is  due to the weak temperature dependence of the minima values for $ \kappa_{xx} $, in contrast to the strong decrease of $ \sigma_{xx} $ with lowering $ T$. 

In summary, the Lorenz number deviates rather strongly from the Wiedemann-Franz value $ L_0 $ as a consequence of the rather complex scattering behavior encountered in our system. This deviation shows further pronounced features through the special scattering properties close to the Lifshitz transition. 

\subsection{Hall resistivity}
The Hall resistivity $ R_{\rm H}= \sigma_{xy}/ B_z  (\sigma_{xx}^2 + \sigma_{xy}^2) $--we set the magnetic field along the $z$ and the current flow along the $x$axis--is used to characterize the nature of the charge carriers. It is interesting to follow the evolution of $ R_{\rm H} $ ($=1/ne $) in the two cases of Fermi surface change. In Fig.~\ref{Fig:BF_R_H} we display the result for $R_{\rm H} $ from the numerical solution of the Boltzmann equations including an out-of-plane magnetic field\cite{buhmann:2013c}. There is a large difference between the Fermi surface tuning by band filling [a)] and uniaxial deformation [b)] considering the magnitude: For the latter tuning, the effect is an order of magnitude larger in the considered range of tuning parameters. The sign change of $ R_{\rm H}, $ Fig.~\ref{Fig:BF_R_H} a), reflects the change from an electron- to a hole-like Fermi surface with a strong temperature dependence of the zero-crossing point. Even at the lowest temperature, however, the sign change of $ R_{\rm H} $ is not exactly at the Lifshitz transition. \

In fact, the current density along the $x$ direction yields the dominant modification of the Fermi distribution along $ [1,0] $. The sign of $R_{\rm H} $ is determined by the curvature of the Fermi surface region, which dominantly carries the current. Looking at Fig.~\ref{Fig:DOS_Gamma_Band_Umklapp} a), the curvature at the Lifshitz transition is convex and yields an electron-like behavior of $ R_{\rm H} $. Only raising the chemical potential $ \mu $ higher leads to emergence of dominantly concave curvature around the $ [1,0] $ direction and thus, to the sign change of $ R_{\rm H} $ at $ \mu \approx 1.25 t $ for $ T = 5 K $. For higher temperature, this feature shifts to lower $ \mu $ due to thermal smearing of the Fermi surface region. 

For the Fermi surface tuned by uniaxial deformation, Fig.~\ref{Fig:BF_R_H} b), the sign of $ R_{\rm H} $ is opposite for positive and negative strain if the temperature is sufficiently high, a behavior which becomes more pronounced with growing $T$. For the lowest temperature displayed, however, $ R_{\rm H} $ is strictly negative. Again, we may consider the curvature of the current-carrying parts for the Fermi surface. For $ \varepsilon_{xx} < 0 $, the Fermi surface around the [1,0] direction remains convex, in other words electron-like, such that $ R_{\rm H} < 0 $. For the other strain direction, the curvature is concave only in a small part near the vHS, which, due to its orientation (with normal vector dominantly along the [0,1] direction), contributes only weakly to the current density. The major contribution to the current density originates from the convex Fermi surface parts with Fermi velocities of sizable $x$ component, as can be anticipated from Fig.~\ref{Fig:DOS_Gamma_Band_Umklapp} b). In this case, however, increasing temperature yields a Fermi surface smearing which yields an effective curvature analogous to the situation we observed in the case of varying chemical potential. The extrema of $ R_{\rm H} $ at the two Lifshitz transitions and their increase are caused by the dips of $ \sigma_{xx} $ which appears quadratically in the denominator of $ R_{\rm H} $ and the strong temperature dependence. 

To our knowledge there are no Hall effect measurements for any of the investigated situation. It would indeed be a helpful test for the quasiparticle picture and our approach to be able to have a comparison with experiments.

\section{Conclusions}
\label{sec:conclusion}
Our analytical and, in particular, numerical investigation of transport properties in a single-band model with a Fermi surface undergoing a Lifshitz transition displays a complex behavior. Our numerical scheme allows us to take the highly anisotropic structure of the electron-electron scattering with Umklapp processes into account. In combination with simple isotropic impurity scattering, we find that the electrical resistivity strongly deviates from the standard Fermi-liquid picture, although our starting point relies on the integrity of the quasiparticle description.

While our study is motivated by experiments on Sr$_{2}$RuO$_{4}$, which possesses three bands at the Fermi level, we focused on a single-band picture including the $ \gamma $ band only. The $ \alpha $  and $ \beta$ bands correspond to hybridized quasi-one-dimensional bands and are only weakly affected by the tuning parameters we have used. In particular, their Fermi surfaces never approach the van Hove points in the BZ. 
Thus, the single-band approximation can be justified on a qualitative level, since the $ \gamma $ band is expected to dominate the anomalous transport properties near the Lifshitz transitions. Nevertheless, in order to obtain a more quantitative picture, all bands should be considered. Moreover, scattering vertex corrections may yield important corrections which have not been taken into account here\cite{buhmann:2013a}. These extensions are referred to future studies.  

While a standard perturbation picture \citep{ziman:1972} of the shifted Fermi surface due to an external electric field fails in the vicinity of the Lifshitz point, our semi-analytical and numerical approach to the solution of the Boltzmann transport equation and the comparison with experiment indicates that a quasiparticle description may indeed be used throughout the whole range of Fermi surface tuning including the Lifshitz points. Analyzing different transport properties, we see that the electron-electron scattering yields strong modifications due to the high anisotropy introduced by Umklapp processes. These processes act in a restricted phase space, which is rather strongly temperature dependent. The considered (non-invasive) Fermi-surface tuning allows to modify this Umklapp phase space and to probe its impact on transport properties. These include not only the electrical conductivity, but also the thermal conductivity, Seebeck coefficient, and in a more indirect way  the Hall effect. 
Our predictions of these general transport properties for the two different Fermi-surface-tuning possibilities allow for further experimental scrutiny of the quasiparticle picture for both charge and heat transport and thus, of the electronic nature of a system close to a van Hove singularity.

\acknowledgements
We would like to thank Clifford Hicks and Andy Mackenzie for many valuable discussions. 
We gratefully acknowledge financial support by the Swiss National Science Foundation through Division II (No. 163186).

\bibliography{ref}

\appendix

\section{Model for the $\gamma$ band of Sr$_{2}$RuO$_{4}$}
\label{app:params}
We approximate the $\gamma$-band, including the effect of uniaxial stress, bu a two-dimensional nearest- and next-nearest-neighbor tight-binding model,
\begin{multline}\label{eq:Dispersion}
	\varepsilon_{\bf k} =-2\left[t_{x}\cos(k_x a)+t_{y}\cos(k_y b)\right]\\
		-2t'\left[\cos(k_x a+k_y b)+\cos(k_x a - k_y b)\right]. 
\end{multline}
Note that in the main text, we use the momentum ${\ve{p}} = \hbar {\bf k}$. For the implementation of the band tuning by uniaxial strain we follow Ref.~\onlinecite{barber:2018} to 
and describe the deformation of the unit cell for the lattice constants $ a $ and $ b$ along $ x $- and $ y $-direction, respectively, by
\begin{align*}
a(\varepsilon_{xx}) = a_0(1+\varepsilon_{xx}),& &b(\varepsilon_{yy}) = a_0(1+\varepsilon_{yy}),
\end{align*}
where $\varepsilon_{xx}$ and $ \varepsilon_{yy} $ are the strain components along the two main axes. For given applied stress $ \sigma_{xx} $ along the $x$-direction the two strain components are connected via the Poisson ratio $ \nu_{xy} $: $ \varepsilon_{yy} = \nu_{xy} \varepsilon_{xx} $ with $ \nu_{xy} < 0 $ and $ \sigma_{xx} = E \varepsilon_{xx} $ ($ E$: Young elasticity modulus).
The lattice deformation modifies of the hopping integrals linearly in the strain,
\begin{align*}
t_x(\varepsilon_{xx}) &= t_0(1-\alpha\varepsilon_{xx}),\quad t_y(\varepsilon_{xx}) = t_0(1+\alpha\nu_{xy}\varepsilon_{xx})\\
t'(\varepsilon_{xx})&=t'_{0}\left(1-\alpha(1-\nu_{xy})\varepsilon_{xx}/2\right),
\end{align*}
where $\alpha$ is a constant parameter adjusting the scale of the effect of the strain.

For numerical calculations, we use the hopping matrix elements $t_0/t = 0.8$, $t'_0/t = 0.3$, the bare chemical potential $\mu_0/t = 1.1$, $\alpha = 10$ and $\nu_{xy}=-0.39$, where $t = 0.14$ eV is the nearest-neighbor-hopping strength of the $\alpha$-band of Sr$_{2}$RuO$_4$ grown on SrTiO$_3$~\cite{hsu:2016}. For the Hall resistivity, we use a magnetic field in the dimensionless units $\mathcal{B}_z = (2\pi)^3 \frac{ea^2}{\hbar}B_z$, with the value $\mathcal{B}_z = 0.1$.

\section{Linearized Boltzmann equation for impurities scattering}
\label{app:impurities}
Here, we derive the linearized Boltzmann equation for small external electric fields ${\ve{E}}$ and temperature gradients ${\ve{\nabla}}_{\ve{r}}T({\ve{r}})$.
\subsection{Collision integral}
First, we calculate the impurity-scattering collision integral Eq.~\eqref{eq:CI} for isotropic point scattering centers, with Eq.~\eqref{eq:CI_Gamma_Fermi}, to lowest order in the correction to the Fermi-Dirac distribution $f_0({\ve{p}})$. Using the expansion of Eq.~\eqref{eq:f_1order} and $f_0({\ve{p}}) = f_0(\varepsilon_{\ve{p}})$, we find
\begin{multline}
	\big[\partial_t f({\ve{p}})\big]_{\rm imp} = -\Omega n_{\rm imp}v_{\rm imp}^2\frac{2\pi}{\hbar}\int (d{\ve{p}'})
	\delta(\varepsilon_{\ve{p}}-\varepsilon_{\ve{p}'})\times \\
	f_0(\varepsilon_{\ve{p}})\big[1 - f_0(\varepsilon_{\ve{p}})\big]\big[\phi({\ve{p}})-\phi({\ve{p}'})\big].
\end{multline}
We can further simplify this expression to
\begin{multline}
	\big[\partial_t f({\ve{p}})\big]_{imp} = -\Omega n_{\rm imp} v_{\rm imp}^2 \frac{2\pi}{\hbar} f_0(\varepsilon_{\ve{p}})\big[1-f_0(\varepsilon_{\ve{p}})\big] \\
	\int_{\varepsilon_{\ve{p}} = \varepsilon_{{\ve{p}}'}} \frac{(d {\ve{p}'})}{|{\ve{\nabla}} \varepsilon_{\ve{p}'}|}\big[\phi({\ve{p}})-\phi({\ve{p}'})\big].
	\label{eq:lin_coll}
\end{multline}
Using the fact that $\phi({\ve{p}}')$ is an odd function of ${\ve{p}'}$ only the first contribution in Eq.~\eqref{eq:lin_coll} survives, and introducing the density of states $N(\varepsilon) = \int_{\varepsilon_{\ve{p}}' = \varepsilon} (d{\ve{p}}')/|{\ve{\nabla}} \varepsilon_{{\ve{p}}'}|$, we find the linearized collision integral
\begin{equation}\label{eq:BE_rhs}
	\big[\partial_t f({\ve{p}})\big]_{imp} =-\underbrace{\frac{2\pi\Omega n_{imp}v_{imp}^2 N(\varepsilon_{\ve{p}})}{\hbar}}_{\tau(\varepsilon_{\ve{p}})^{-1}}f_0(\varepsilon_{\ve{p}})\big(1-f_0(\varepsilon_{\ve{p}})\big)\phi({\ve{p}}).
\end{equation}
As expected for point scattering centers the scattering time is angle independent (s-wave scattering).

\subsection{Temperature gradient and external electric field}
Introducing ${\ve{v}}_{\ve{p}} \equiv \dot{\ve{r}}= {\ve{\nabla}}_{\ve{p}} \varepsilon_{\ve{p}}$, we can write the second term of the left-hand side of Eq.~\eqref{eq:BE_full} to lowest order as
\begin{eqnarray}\label{eq:Tgrad}
	\dot{{\ve{r}}} \cdot {\ve{\nabla}}_{\ve{r}}f({\ve{p}}) &=&{\ve{v}}_{\ve{p}}  \cdot {\ve{\nabla}}_{\ve{r}}f({\ve{u}})\\
		&=& f_0(\varepsilon_{\ve{p}})\big[1-f_0(\varepsilon_{\ve{p}})\big]\left(\frac{\varepsilon_{\ve{p}}-\mu}{T^2}\right){\ve{\nabla}}_{\ve{r}}T({\ve{r}})\cdot {\ve{v}}_{\ve{p}}.
\end{eqnarray}

Further, we calculate the third term on the left-hand side of Eq.~\eqref{eq:BE_full}. Considering only an electric field, i.e., $\dot{\ve{p}} = -e {{\ve{E}}}$, this yields to lowest order in the applied field ${\ve{E}}$
\begin{eqnarray}
	\dot{{\ve{p}}}  \cdot{\ve{\nabla}}_{\ve{p}}f({\ve{p}}) &=& -e {{\ve{E}}}  \cdot {\ve{\nabla}}_{\ve{p}}f({\ve{p}})\\
	 &=& -e\frac{{{\ve{E}}}\cdot{\ve{v}_{\ve{p}}}}{T} f_0({\ve{p}}) \big[1-f_0({\ve{p}})\big].
	 \label{eq:el}
\end{eqnarray}

Using Eq.~\eqref{eq:BE_rhs}, \eqref{eq:el} and \eqref{eq:Tgrad} we find the linearized Boltzmann equation
\begin{equation}\label{eq:lin_comp}
	\left[\partial_t + \tau(\varepsilon_{\ve{p}})^{-1} \right]\phi({\ve{u}})=-\left[\left(\frac{\varepsilon_{\ve{p}}-\mu}{T}\right){\ve{\nabla}}_{\ve{r}}T({\ve{r}}) + e{{\ve{E}}}\right]\cdot\frac{{\ve{v}}_{\ve{p}}}{T},
\end{equation}
which finally leads to the stationary solution
\begin{equation}\label{eq:lin_stat_app}
	\phi({\ve{p}})=-\left[\left(\frac{\varepsilon_{\ve{p}}-\mu}{T}\right){\ve{\nabla}}_{\ve{r}}T({\ve{r}}) + e{{\ve{E}}}\right]\cdot\frac{{\ve{v}}_{\ve{p}}}{T}\tau(\varepsilon_{\ve{p}}).
\end{equation}

\end{document}